\documentclass[11pt]{article}
\usepackage{hyperref}
\pdfoutput=1 
\begin{document}

\title{Flows in inkjet-printed aqueous rivulets}

\author{Vadim Bromberg, Timothy J. Singler \\
 \\
\vspace{6pt}
 Mechanical Engineering Department, \\
 State University of New York at Binghamton, Binghamton, NY 13902,
USA}

\maketitle
\begin{abstract}
We used optical microscopy to investigate flows inside water rivulets that
were inkjet-printed onto different surfaces and under different ambient
conditions. The acquired fluid dynamics videos were submitted to the 2013 Gallery
of Fluid Motion. 
\end{abstract}

\section{Introduction}

Understanding the flow inside sessile liquid structures of different shapes is important in a variety of solution-based material deposition
and patterning processes. Solvent evaporation inherent in these processes
is already known to lead to a rich variety of flows \cite{Deegan1997}.
The small length scale and general lack of shape symmetry implies
the potential for capillary pressure gradients and corresponding flow
phenomena. Finally, the non-instantaneous nature of the formation
of these liquid structures adds another element to the flow
complexity.

In the linked videos, we investigated the internal flow of inkjet-printed
water rivulets of finite length using optical microscopy. Six millimeter-long
rivulets were formed by printing a pre-determined number of drops
($D_{0}=52\mu$m) at controlled frequency ($f=20$Hz)
and spatial overlap ($\Delta x=20\mu$m). Microscope cover
slips made of borosilicate glass were surface coated and used as substrates.
Two surface coatings were investigated - S1805 photoresist with and
without KOH etching. Rivulets were printed inside a controlled humidity
chamber at two relative humidity levels - $45\%$ and $25\%$. The
ambient, ink, and substrate were kept at room temperature ($T=25$C).
For fluorescent microscopy videos, the water was seeded with Nile
Red polystyrene spheres ($1.1\mu$m) at $0.1\%$ volume fraction.
We report the experimental details and results of a wider range of
both printing parameters, ink properties, and substrate surfaces elsewhere
\cite{BrombergDFD2013}. 

The two surface coatings resulted in different values of the static
advancing and receding contact angles ($\theta_{A}$ and $\theta_{R}$)
for water and in the rivulet formation process during printing. Water
droplets on the photoresist-coated surface exhibited $\theta_{A}=80^{\circ}$
and $\theta_{R}=35^{\circ}$while those on the etched surface $\theta_{A}=45^{\circ}$
and $\theta_{R}=0$. Complete rivulet formation was inhibited on the
first surface type due to a non-zero $\theta_{R}$, which allowed
capillary-driven contact line recession and rivulet breakup into individual
droplets \cite{Schiaffino1997}. The zero $\theta_{R}$ on the second
surface type prevented rivulet breakup \cite{Schiaffino1997} but
resulted in the formation of a distinctive bulge at the starting end
of the water rivulet. The bulge grows immediately after the coalescence
of the first few drops during rivulet formation. 

We investigated the flow that causes bulge growth
by using optical fluorescent microscopy. During rivulet formation,
a distinctly pulsatile axial flow drives fluid away from the terminal
end of the growing rivulet where printed drops are landing. The frequency of the flow is the same as the drop frequency. It has been hypothesized that
the large local mean curvature in the region where drops land causes a capillary pressure gradient
along the length of the rivulet \cite{Duineveld2003}. Using
$\mu$PIV, the height- and width-averaged axial velocity was measured
and showed that the amplitude of the pulsatile flow decreases with
reduced relative humidity in the ambient, all other conditions being
fixed. The decreased flow magnitude results in the inhibition of
bulge formation. 

\section{Videos}
The video entry to the 2013 Gallery of Fluid motion is shown in
\begin{itemize}
\item \href{URL of video}{Video 1 - Low resolution}
\item \href{URL of video}{Video 2 - High resolution}
\end{itemize}
In the video, the following subjects are shown:

\begin{itemize}
\item A schematic of the flow observation approach
\item Visualization of rivulet break up during printing onto a substrate with non-zero receding contact angle (surface type 1)
\item Visualization of successful rivulet formation on surface type 2 and concequent bulge growth
\item Fluorescence microscopy video showing pulsatile axial flow and corresponding mean axial speed
\item The effect of liquid evaporation on axial flow and bulge growth inhibition 

\end{itemize}

\end{document}